\documentclass[twocolumn]{aa}
\usepackage{natbib}
\usepackage{color}
\usepackage{ragged2e}
\usepackage[pdfpagelabels=false]{hyperref}	
\hypersetup{colorlinks=true,linkcolor=blue,citecolor=blue,filecolor=blue,urlcolor=blue,}
\usepackage[varg]{txfonts}
\usepackage{graphicx,rotating}
\usepackage[normalem]{ulem}
\usepackage{colortbl}
\usepackage{xcolor}
\usepackage{bbold}

\bibpunct{(}{)}{;}{a}{}{,} 

\definecolor{darkolivegreen}{rgb}{0.33, 0.42, 0.18}
\definecolor{salmon}{rgb}{0.95,0.5,0.25}

\newcommand{\rhm}{r_\text{hm}}

\newcommand{\kpc}{\mathrm{\, kpc}}

\newcommand{\mFDM}{m_{22}}

\newcommand{\Ccdm}{\mathcal{C}_\text{CDM}}

\begin{document}

\title{Fuzzy dark matter dynamical friction:}
\subtitle{stalling of globular clusters induced by dynamical heating}

\titlerunning{Fuzzy dark matter dynamical friction}
\author{Adrian Szpilfidel \inst{1}, Pierre Boldrini \inst{1}, Jo Bovy \inst{2} and Paola Di Matteo \inst{1}}

\offprints{Adrian Szpilfidel, \email{adrian.szpilfidel@obspm.fr }}
\institute{$^{1}$ LIRA, Observatoire de Paris, Université PSL, Sorbonne Université, Université Paris Cité, CY Cergy Paris Université, CNRS, 92190 Meudon, France, \\$^{2}$ David A. Dunlap Department of Astronomy and Astrophysics, University of Toronto, 50 St George Street, Toronto, ON M5S 3H4, Canada}
\authorrunning{Szpilfidel et al.}
\date{submitted to A$\&$A}

\abstract{We present a new implementation of fuzzy dark matter (FDM) dynamical friction within the \texttt{galpy} framework, enabling orbital integrations of globular clusters (GCs) across a broad range of halo-to-GC mass ratios and boson masses. In this alternative DM scenario, dynamical friction is reduced or even suppressed by heating induced by FDM density granules. We further quantify the role of baryons and solitonic cores, natural consequences of FDM in galaxies, on the efficiency of orbital decay and the long-term survival of GCs. The most significant deviations from the cold DM (CDM) paradigm arise in the dwarf-galaxy regime, where FDM dynamical friction can stall the inspiral of GCs over a Hubble time, thereby preventing their sinking into galactic centers and halting the canonical galactic cannibalism of clusters. Importantly, our FDM-only friction model should be regarded as a conservative lower bound, since the inclusion of realistic FDM cores can only strengthen the survival of GCs through core stalling. This stalling mechanism not only preserves in-situ populations that would otherwise be erased in CDM, but also strongly limits the mixing of in-situ and ex-situ clusters, yielding a bimodal radial distribution of GCs. Our results show that the demographics of GC systems encode a distinct dynamical signature of FDM in dwarfs. These predictions open a new pathway to constrain the boson mass parameter with upcoming Euclid DR1 observations of extragalactic GCs, while simultaneously offering a natural explanation for the long-standing Fornax timing problem.}

\maketitle



\section{Introduction}

Dynamical friction (DF) is a fundamental mechanism of energy loss in galactic dynamics. It plays a crucial role in the evolution of massive objects such as star clusters, black holes, satellite galaxies, or globular clusters (GCs) \citep{2008gady.book.....B}. Consider a massive object of mass $M_{\rm obj}$ orbiting within a gravitational potential generated by a background of particles of dark matter (DM) or stars with individual mass $m_{\rm p}$. In the regime where $M_{\rm obj} \gg m_{\rm p}$, such as a GC ($M_{\rm obj} \sim 10^6\, \mathrm{M}_\odot$) moving through a galaxy, the object gravitationally deflects nearby background particles. Locally, the gravitational pull of the object dominates over that of the surrounding potential, creating an overdensity or gravitational wake behind it. This asymmetry induces a net gravitational force opposing the object’s motion, leading to deceleration and orbital energy loss. Over time, the object spirals inward toward the galactic center. This process is known as DF \citep{Chandra43}.

Since DF directly depends on the distribution of background particles within the galaxy, its behavior can serve as an indirect probe of DM properties. It is therefore legitimate to investigate how changes in the nature of DM may affect the strength and character of DF. Among various alternative DM models, fuzzy DM (FDM) has received increasing attention for addressing several shortcomings of the standard cold DM (CDM) paradigm, particularly on small galactic scales \citep{2017ARA&A..55..343B,2021Galax..10....5B}. In this framework, DM is modeled as a non-relativistic, ultra-light scalar field with no self-interaction \citep{2000NewA....5..103G,2000PhRvL..85.1158H}. The constituent particles of FDM have extremely low masses, typically in the range $m_\chi = 0.1$--$100 \times 10^{-22}$~eV. Such low masses give rise to quantum mechanical effects on astrophysical scales. A key quantity characterizing these quantum effects is the de Broglie wavelength, $\lambda_B = \hbar / (m_\chi v)$, which defines the scale below which wave-like behavior dominates. For small FDM particle mass, $\lambda_B$ can reach kpc scales, leading to significant modifications of the DM density distribution on galactic scales. This results in distinctive dynamical features: the suppression of low-mass subhalo formation \citep{Marsh14,Mocz17,Chiang21,2021ApJ...916...27D,Schive16,May23,2022MNRAS.511..943C}, the emergence of solitonic cores with approximately constant central density \cite{Schive14}, and a dynamical heating caused by fluctuations in the gravitational potential \citep{Hui17,BO19,EZ20,2021ApJ...916...27D}. In this alternative DM model, these fluctuations mainly arise from a turbulent density field within FDM halos, characterized by the presence of density granules. Unlike the stochastic and rare perturbations caused by subhalos in CDM, FDM fluctuations are continuous and recurrent, leading to a diffusive heating process. For instance, this FDM-induced heating can cause the thickening of cold stellar streams from dissolving GCs in the Milky Way (MW) \citep{Amorisco18}, or it may explain the observed thickening of the Galactic disk over a Hubble time \citep{Chiang23,2024arXiv241213275H}. More importantly, this distinctive feature of FDM can lead to a reduction, or even a suppression, of DF compared to the CDM case \citep{Hui17,BO19,Lancaster20}. We discuss this key aspect further in Section \ref{explication}.

This last point has important implications. The reduced efficiency of DF in FDM core galaxy has been proposed as a mechanism to stall the orbital decay of GCs and black holes, potentially allowing them to survive in galaxies over a Hubble time \citep{Hui17,BO19}. The modeling of DF in the FDM context has been progressively developed by \cite{Hui17}, \cite{BO19}, and \cite{Lancaster20}. All theses studies rely on the Madelung formalism to describe FDM as a fluid. This approach is crucial, as it allows the wave-like nature of FDM to be treated within a hydrodynamical framework. \cite{Lancaster20} provided a detailed exploration of DF in FDM, combining analytical treatments (for point masses, extended mass distributions, and velocity-dispersed backgrounds) with fully non-linear numerical simulations.

GCs are excellent tracers of DF for several reasons. These gravitationally bound stellar systems are both compact and relatively massive ($10^{4}$--$10^{6}$ M$_{\odot}$), which maximizes the efficiency of this drag force. GCs also span a wide range of galactocentric distances, and their diversity of orbits enables us to probe how DF operates as a function of the local density. Finally, GCs are found in virtually all types of galaxies, across a wide range of masses \citep{2025ApJ...978...33L}, allowing us to investigate the behavior of this force in different environments. From an observational perspective, the third data release of the Gaia mission provides full 6D phase-space information for all Galactic GCs \citep{Gaia21,2021MNRAS.505.5978V}, while Euclid is expected to revolutionize the number of detections in nearby galaxies thanks to its unprecedented coverage of galaxies spanning stellar masses from $10^{9}$ (dwarfs) to $10^{12}$ M$_{\odot}$ (MW-like systems) \citep{Voggel25}. This makes GCs ideal natural probes for studying the efficiency of DF and its dependence on DM properties.

In this paper, we implement the FDM DF formalism of \cite{Lancaster20} into the \texttt{galpy} library\footnote{\url{https://github.com/jobovy/galpy}} \citep{Bovy15}, with the goal of exploring, through orbital integrations, the subtle differences that arise from (i) the suppression of DF and (ii) the presence of a DM core, in shaping the long-term dynamics of GCs. These clusters are modeled as point-mass tracers of the gravitational potential, allowing us to isolate and quantify the individual contributions of these two key features of FDM. The paper is structured as follows: Section 2 introduces the theoretical formulation of FDM DF and its implementation in the \texttt{galpy} framework, together with tests of validity and an application to the Fornax GC system. In Section 3, we present and discuss our results, focusing on the efficiency of FDM DF across different halo masses, the impact of baryons and solitonic cores, and the resulting demographics of in-situ and ex-situ GC populations. Finally, Section 4 summarizes our main conclusions.

\section{Fuzzy dark matter dynamical friction}

\subsection{Formulation}

\begin{figure}[t]
    \centering
    \includegraphics[width=\linewidth]{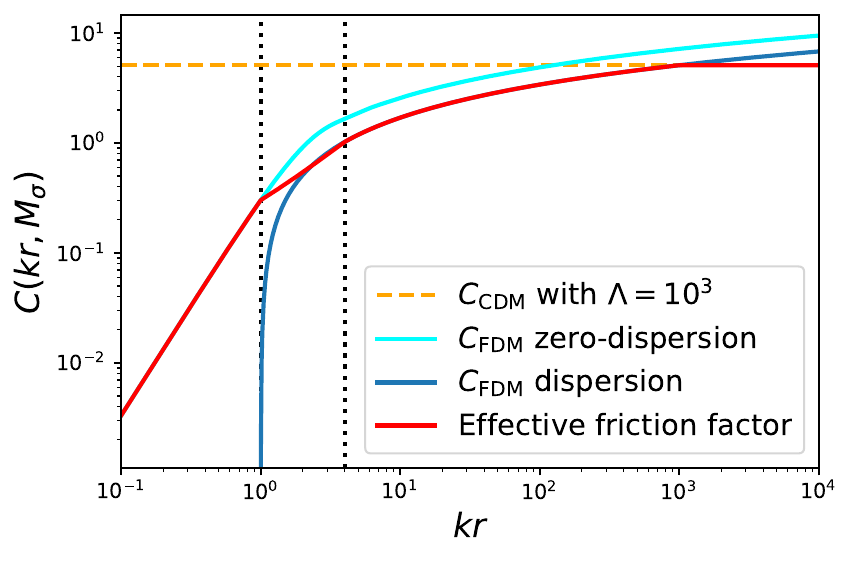}
    \caption{Computed FDM coefficient $\mathcal{C}_{\rm FDM}$ as function of the dimensionless parameter $kr$ for different regimes, assuming a fixed $M_\sigma=2$. Orange dashed line represents the classical coefficient $\Ccdm$, calculated with fixed $\Lambda=10^3$. The black dashed vertical lines mark $M_\sigma/2$ and $2M_\sigma$.}
    \label{fig:Crel}
\end{figure}

The classical formulation of the DF force was derived by \cite{Chandra43}, assuming a Maxwellian velocity distribution for the background particles:
\begin{equation}
\label{eq:ChandraDF}
    \vec{F}_{\rm CDM} = -4\pi G^2 M_{\rm GC}^2 \, \rho(r) \, \frac{\vec{v}}{v^3} \, \mathcal{C}_{\rm CDM}(r,v) \quad ,
\end{equation}
where the coefficent $\mathcal{C}_{\rm CDM}(r,v)$ is given by:
\begin{equation}
\label{eq:Ccl}
    \mathcal{C}_{\rm CDM}(r, v) = \frac{1}{2} \ln{(1+\Lambda^2)}\left[ \mathrm{erf}\left(\frac{v}{\sqrt{2} \sigma}\right) - \frac{2v}{\sqrt{2\pi} \sigma} \, e^{-\frac{v^2}{2\sigma^2}} \right],
\end{equation}
and the Coulomb factor $\Lambda$ is expressed as:
\begin{equation}
\label{eq:lnLambda}
\Lambda= \frac{r}{\max(r_{\mathrm{hm}},\, b_{\mathrm{min}})},
\end{equation}
where $r$ is the galactrocentic distance of the GC, $v$ its velocity, $M_{\rm GC}$ its mass, and $\rho(r)$ the local density of the host galaxy. Our dynamical friction implementation follows \cite{Petts15}, which introduced a semi-analytic model based on \cite{Chandra43} formalism, incorporating radially varying minimum and maximum impact parameters. This approach provides excellent agreement with full N-body simulations for both cuspy and core DM distributions without any parameter fine-tuning and successfully reproduces the core stalling effect. The CDM coefficient defined in Equation~\eqref{eq:Ccl} is shown as a dashed orange line in Figure~\ref{fig:Crel}, for a fixed value of $\Lambda = 10^{3}$. The parameter $b_{\mathrm{min}} = G M_{\rm GC} / v^2$ corresponds to the impact parameter leading to a $90^\circ$ deflection, while $r_{\mathrm{hm}}$ is the half-mass radius of the cluster (typically about 10 pc). $G$ is the gravitational constant. The formalism of \cite{Chandra43} was originally derived under the assumption of an infinite, homogeneous background. Despite this limitation, our implementation accounts for the finite size of GCs by incorporating their half-mass radius, $r_{\rm hm}$, into the Coulomb logarithm. In addition, its effectiveness can be understood because dynamical friction in real systems arises from resonant interactions that, when forming an effective continuum, recover Chandrasekhar’s expression \citep{Tremaine84,Weinberg86}. Indeed, \cite{Petts15} implementation has been extensively validated against $N$-body simulations, successfully reproducing the inspiral of GCs in spherical halos without parameter fine-tuning. We therefore apply the formulation locally along the orbit, while noting that its validity is expected to hold for approximately spherical systems, although extensions to strongly non-spherical environments, such as discs, remain uncertain. The DF force is proportional to the mass squared of the orbiting GC and to the local density of the host galaxy, implying that DF is more efficient for a massive GC moving through a dense environment. Moreover, the characteristic timescale over which the GC's apocentre is significantly reduced by DF is approximately given by $t_{\mathrm{fric}} \sim \left( M_{\mathrm{gal}}(r) / M_{\mathrm{GC}} \right) \times t_{\mathrm{dyn}}$, where $M_{\mathrm{gal}}(r)$ is the mass of the host galaxy enclosed at GC radius $r$, and $t_{\mathrm{dyn}}$ is the orbital time \citep{2008gady.book.....B,DynamicsAstrophysicsGalaxies}. For MW-like galaxies, the enclosed mass is typically much larger than the mass of a GC, rendering DF inefficient on a Hubble timescale. Additionally, inspecting the Coulomb logarithm, particularly its dependence on $1/r_{\mathrm{hm}}$, shows that DF is less effective for extended GCs: their mass is more spatially diffuse, which results in a less dense gravitational wake.

\begin{figure*}
    \centering
    \includegraphics[width=1\linewidth]{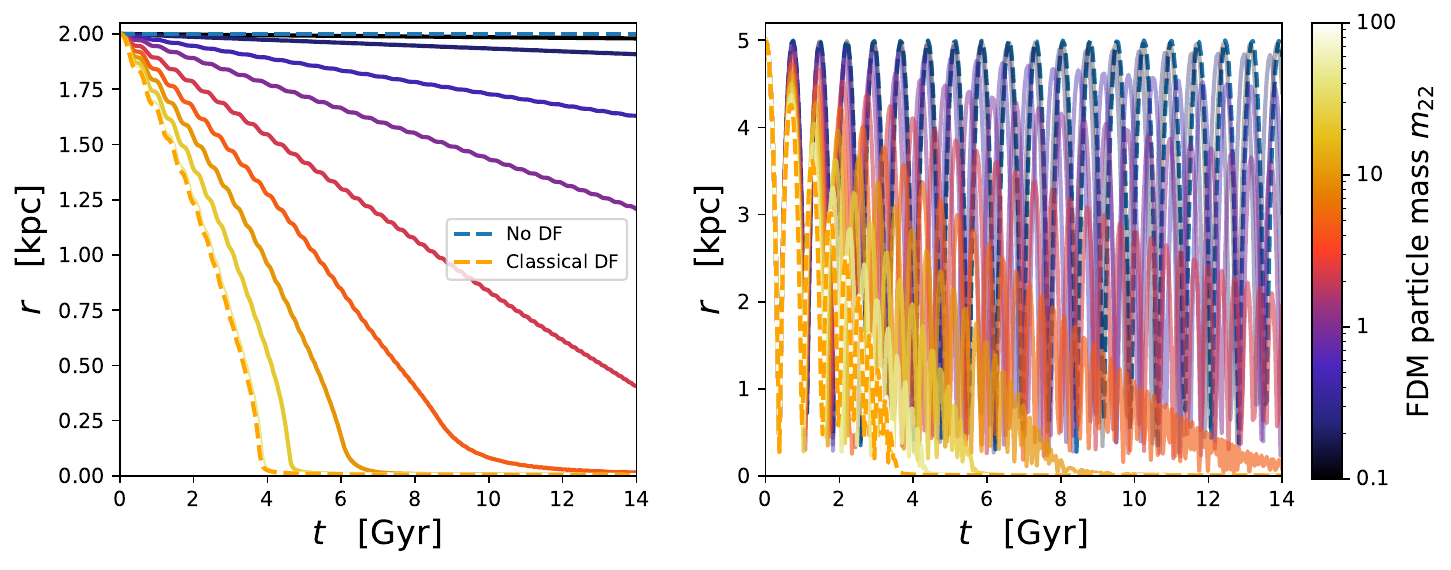}
    \caption{FDM dynamical friction: Orbital radius as a function of time for a $10^6$ M$_\odot$ GC on circular (left panel) and radial (right panel) orbits within a static NFW halo, shown for different values of the FDM particle mass $m_{22}$. The dashed blue curve shows the evolution in the absence of any DF, while the dashed orange curve corresponds to the classical Chandrasekhar friction.}
    \label{figure2}
\end{figure*}

The DF experienced by a GC infalling into a FDM galaxy differs significantly from the classical formulation, which applies to CDM. This difference arises from the wave-like nature of FDM particles. Quantum interference between FDM waves throughout the halo generates persistent and stochastic density fluctuations known as granules. These fluctuations create localized over- and under-densities due to constructive and destructive interference, respectively. The typical size of these granules is set by the de Broglie wavelength, $\lambda_B$. When the characteristic size of the GC, $l$, satisfies $l \ll \lambda_B$, the DF is modified by oscillations in the density of the gravitational wake. In this regime, the frictional force is reduced compared to the classical Chandrasekhar estimate. Conversely, for $l \gg \lambda_B$, the classical CDM regime is recovered. A useful diagnostic to distinguish between these regimes in an FDM context is the quantum Mach number~\citep{Lancaster20}:
\begin{equation}
    \mathcal{M}_Q = 44.56 \left( \frac{v}{1\, \mathrm{km\, s}^{-1}} \right) 
\left( \frac{M_{\mathrm{GC}}}{10^5 \mathrm{M}_\odot} \right)^{-1} m_{22}^{-1},
\label{MQN}
\end{equation}
where $m_{22} = m_{\chi} / 10^{-22}\, \mathrm{eV}$, and $m_\chi$ is the FDM particle mass. In the regime where $\mathcal{M}_Q \gg 1$, the DF force is well approximated by~\citep{Hui17,Lancaster20,BO19}:
\begin{equation}
    \label{eq:FDMDF}
    \vec{F}_{\rm FDM} = -4\pi G^2 M_{\rm GC}^2 \, \rho(r) \, \frac{\vec{v}}{v^3} \, \mathcal{C}_{\rm FDM}(kr, M_{\sigma}) \quad ,
\end{equation}
which resembles the classical expression (see \autoref{eq:ChandraDF}), but with a modified coefficient $\mathcal{C}_{\rm FDM}(kr, M_{\sigma})$ that depends on two dimensionless parameters. The first is the wave number, defined as:     
\begin{equation}
\label{k}
    k = \frac{m_{\chi} v}{\hbar} \quad ,
\end{equation}
where $\hbar$ is the reduced Planck constant. 

Therefore, the amplitude of these deviations from the classical CDM DF depends directly on the free parameter of FDM theory: the mass of the ultralight boson $m_{\chi}$. The second is the classical Mach number, given by: 
\begin{equation}
\label{Msigma}
    M_{\sigma} = \frac{v}{\sigma} \quad ,
\end{equation}
where $\sigma$ is the local velocity dispersion of the host galaxy.

The introduction of the classical Mach number, $M_\sigma = v / \sigma$, allows one to distinguish between FDM halos with negligible velocity dispersion (static backgrounds) and those with velocity-dispersed backgrounds. In the limit $M_\sigma \gg kr$, the FDM medium can be approximated as having no velocity dispersion, i.e., as a static background with constant density. In this zero-velocity dispersion regime, the DF coefficient simplifies to \citep{Hui17,Lancaster20}:
\begin{equation}
\label{0-disp}
    \mathcal{C}_{\rm FDM}(kr) = \mathrm{Cin}(2kr) + \frac{\sin(2kr)}{2kr} - 1,
\end{equation}
where
\begin{equation}
    \mathrm{Cin}(z) = \int_0^z \frac{1 - \cos(t)}{t} \, \mathrm{d}t
\end{equation}
is the cosine integral function. The FDM coefficient with zero-dispersion defined in Equation~\eqref{0-disp} is shown as a cyan line in Figure~\ref{fig:Crel}. However, in more realistic scenarios, the FDM medium is expected to exhibit a non-negligible velocity dispersion, corresponding to the regime $M_\sigma \ll kr$. In this case, the inhomogeneities induced by the FDM waves are smoothed out over scales comparable to or smaller than the system size, and the friction coefficient becomes ~\citep{BO19,Lancaster20}:
\begin{equation}
    \mathcal{C}_{\rm FDM}(kr, M_{\sigma}) = \ln\left( \frac{2kr}{M_\sigma} \right) 
    \left[ \mathrm{erf}\left( \frac{v}{\sqrt{2} \sigma} \right) - \frac{2v}{\sqrt{2\pi} \sigma} e^{- \frac{v^2}{2\sigma^2}} \right].
    \label{disp}
\end{equation}

The FDM coefficient with dispersion defined in Equation~\eqref{0-disp} is shown as a blue line in Figure~\ref{fig:Crel}, for a fixed value of $M_{\sigma} = 2$. Our implementation of the \cite{Lancaster20} formulation for FDM DF is detailed in Section~\ref{A1}, and corresponding \texttt{galpy} documentation is already available\footnote{\url{https://docs.galpy.org/en/latest/reference/potentialfdmdynfric.html}}. We have also performed tests to ensure the validity of our new implementations, which are described in Section~\ref{A2}. This formulation assumes that the perturber moves linearly through a collisionless medium, as originally derived in the pioneering work of \cite{Chandra43}. More recently, DF has been derived by \cite{2023PhRvD.107b3516B} for perturbers on circular orbits in FDM backgrounds. Applying this framework to the Fornax GCs, they showed that for clusters located within (outside) the scale radius of Fornax ($r_s = 0.8$ kpc), the classical FDM DF prescription tends to underestimate (overestimate) the strength of dynamical friction. We therefore note that the use of the classical Chandrasekhar formulation in the FDM case should be regarded as a caveat of our study.

\subsection{Why dynamical friction is reduced in FDM galaxies?}
\label{explication}

DF in an FDM universe does not intrinsically decay in the sense of a vanishing frictional force, but its effect on an infalling GC can diminish or even vanish as it approaches the galactic center. This behavior is depicted in Figure~\ref{figure2}, and results from the persistent stochastic density fluctuations of the FDM halo, which dynamically heat the GC. This heating process, specific to the wave-like nature of FDM, counteracts the energy loss due to DF by injecting kinetic energy into the cluster. These FDM fluctuations can be modeled as effective "quasiparticles" of mass $m_{\mathrm{eff}}$, representing FDM granules \citep{BO19}:
\begin{equation}
    m_{\mathrm{eff}} = \frac{\pi^{3/2} \hbar^3 \rho(r)}{m_\chi^3 \sigma^3} \quad ,
\end{equation}
where $\rho(r)$ is the local DM density. Notably, $m_{\mathrm{eff}}$ scales inversely with $m_\chi$. As a GC spirals toward the center of an FDM halo, the local density $\rho(r)$ increases, causing $m_{\mathrm{eff}}$ to grow substantially. When the GC mass $M_{\mathrm{GC}}$ becomes comparable to twice the effective mass of the FDM quasiparticles ($M_{\mathrm{GC}} \sim 2 m_{\mathrm{eff}}$), the energy injected into the GC by the stochastic fluctuations becomes comparable to the energy lost through DF \citep{BO19}. At this point, the inspiral effectively stalls: heating and DF balance out, and the net effect of DF drops to nearly zero. In our modeling, this FDM heating is included in  in the FDM dynamical friction coefficients described by Equations~\eqref{0-disp} and~\eqref{disp}.

\begin{table}
\centering
\setlength{\tabcolsep}{3.6pt}
\caption{Fornax globular clusters}
\begin{tabular}{|c|c|cc|ccc|}
\hline
 &  & \multicolumn{2}{|c|}{CDM $\tau$} & \multicolumn{3}{|c|}{FDM $\tau$} \\
\hline
 &  & Hui+17 & Here & Hui+17 & Here & Here \\
$R$ & $M_{\text{GC}}$ & - & - & $m_{22} = 3$ & 3 & 20 \\
\multicolumn{1}{|c|}{[kpc]} &
\multicolumn{1}{c|}{$[10^5\,M_\odot]$} &
\multicolumn{2}{c|}{[Gyr]} &
\multicolumn{3}{c|}{[Gyr]} \\
\hline
1.85 & 0.37 & 112  & 25   & 215 & 393 & 176 \\
1.21 & 1.82 & 9.7  & 2.56 & 12  & 59  & 26 \\
0.50 & 3.63 & 0.62 & 0.30 & 2.2 & 19  & 11 \\
0.28 & 1.32 & 0.37 & 0.24 & 10  & 24  & 13 \\
1.65 & 1.78 & 21.3 & 5.08 & 31  & 74  & 34 \\
\hline
\end{tabular}
\parbox{\hsize}{\small Notes: From left to right, the columns report the observed projected distances and masses of the five GCs, followed by a comparison between the infall times $\tau$ computed in this work and those from \citet{Hui17}, under both CDM and FDM assumptions for the Fornax halo. We have tested two fixed FDM particle masses of $m_{22} = 3$ and 20. The infall times $\tau$ correspond to the time for the GC to fall below $0.14\kpc$, which corresponds to 10\% of the core radius of the FDM halo.}
\label{Table1}
\end{table}

Figure~\ref{figure2} illustrates the dependence of FDM DF on the FDM particle mass $m_{22}$, for both circular (left panel) and radial (right panel) orbits in a static NFW halo. In the left panel of Figure~\ref{figure2}, for small values of $m_{22}$ (more wave-like behavior), the friction is significantly suppressed and the GC remains at large radius for more than a Hubble time. As $m_{22}$ increases, the FDM behavior approaches the classical CDM regime. The granules become smaller, the quantum heating effect weakens, and the gravitational wake becomes more efficient. This leads to a progressively shorter infalling time, and the orbits gradually converge toward the classical friction result. At high $m_{22}$, the GC reaches the center in 10 Gyr, similar to the dashed orange curve (see Figure~\ref{figure2}). The right panel of Figure~\ref{figure2} shows the same analysis for a radial orbit with a similar qualitative effect of $m_{22}$. Low $m_{22}$ values lead to prolonged survival and incomplete decay within a Hubble time, while high $m_{22}$ values allow the orbit to decay rapidly, with the pericentre shrinking progressively with each oscillation. Starting from intermediate values of the FDM particle mass (e.g., $m_{22} \sim 0.5$), the apocentres of the orbit begin to exhibit a clear plateau. In this regime, the orbital decay gradually slows down and eventually stalls as the cluster approaches the center. This plateau indicates that the energy lost through DF is effectively balanced by dynamical heating from FDM granules, thereby preventing further infall. Figure~\ref{figure2} clearly demonstrates the strong dependence of the effectiveness of FDM DF on the particle mass. For low $m_{22}$ (strong wave regime), the combination of dynamical heating can completely halt the infall. In contrast, for large $m_{22}$ values, the FDM model reproduces classical CDM-like friction. These results highlight the central role of $m_{22}$ in determining the long-term survival of orbiting objects in FDM halos, and provide a potential observational handle to constrain the FDM parameter space. 

Quantum interference in the FDM gravitational potential also modifies the central density profile of DM halos. Instead of forming a steep central cusp, as described by the NFW profile in CDM, FDM halos develop a flat core. This structural difference has two major consequences on the DF undergone by GCs. First, the presence of a central core increases the mass of the FDM quasiparticles. This is because in the central regions of a cored halo, the ratio $\rho / \sigma^3$ typically scales as $1/r^3$, whereas for a cuspy NFW profile, it behaves as $\rho / \sigma^3 \sim 1/r^{5/2}$. Since the effective mass of quasiparticles scales as $m_{\mathrm{eff}} \propto \rho / \sigma^3$, this results in more massive quasiparticles in the core-dominated regime, thereby enhancing the efficiency of dynamical heating. As a result, heating becomes more effective at counteracting DF in the presence of a core. Second, independently of this heating process, DF itself is directly reduced in the core because the force depends linearly on the local density $\rho(r)$, as seen in Equations~\eqref{eq:ChandraDF} and~\eqref{eq:FDMDF}. The shallower central density profile in FDM halos leads to a weaker gravitational wake and, consequently, a diminished frictional force. Thus, both the enhanced heating and the suppressed friction contribute to the stalling of GC inspiral in the central regions of FDM halos.

To model an FDM halo, we combined a central solitonic core with an outer power-law envelope. The final FDM profile is thus constructed as the sum of two components using the \texttt{TwoPowerSphericalPotential} profile available in \texttt{galpy}: a solitonic profile approximated by a power law:
\begin{equation}
\rho(r) = \frac{\rho_{01}}{4\pi r_{c1}^3}\left[ 1 + \left( \frac{r}{r_{c1}} \right) \right]^{-\beta},
\label{galpy1}
\end{equation}
and an outer pseudo-NFW envelope, ensuring a physical transition and matching the CDM halo outside the core region, approximated by:
\begin{equation}
\rho(r) = \frac{\rho_{02}}{4\pi r_{c2}^3}\left[ 1 + \left( \frac{r}{r_{c2}} \right) \right]^{-3}.
\label{galpy2}
\end{equation}
We optimized the parameters $(\rho_{01}, \rho_{02}, r_{c1}, r_{c2})$ by minimizing the quadratic error to reproduce the combined soliton + NFW envelope density. Our profile combination ensures a transition between the core and the NFW envelope with a slope of $-1$ as predicted by FDM. Ultimately, our FDM potential is approximated by a sum of existing \texttt{galpy} potentials, which offers a good compromise between accuracy and performance since they are faster than a newly implemented potential not yet optimized for \texttt{galpy} .

\subsection{Application to Fornax globular clusters}

In order to test our new \texttt{Python} class \texttt{FDMDynamicalFrictionForce}, we apply it to the case of the Fornax dwarf spheroidal galaxy and its GCs. A well-known paradox surrounding the Fornax system is that, under standard CDM assumptions, DF should have caused all five GCs to spiral into the galaxy's center within a Hubble time \citep{2000ApJ...531..727O}. Yet, all of them are still observed orbiting even in the dwarf central region. One way to alleviate this discrepancy is to consider alternative DM models, such as FDM, which modify the DF acting on the GCs and can significantly extend their survival times. We have computed the infalling time $\tau$ for the five Fornax GCs evolving in a static CDM and FDM halo, by numerically integrating their orbits using \texttt{galpy}. For the CDM halo, we adopt a NFW profile with a scale radius of $r_s = 0.8\,\mathrm{kpc}$ and a halo mass of $1.98 \times 10^8$ M$_\odot$. For the FDM model, we use a cored profile with a larger core of $r_c = 1.4\,\mathrm{kpc}$ and halo mass $8 \times 10^8$ M$_\odot$ (see Table 2 of \cite{2012MNRAS.426..601C}). For the Fornax case, we define the infalling time as the time required for the apocentre of the GC’s orbit to fall below 10\% of the DM core radius. This contrasts with the approach taken by \citet{Hui17}, who estimated falling times by solving the orbital decay equation using constant friction coefficients. The initial conditions for the GCs in our analysis are set by their currently observed masses and projected galactocentric positions, assuming circular orbits. All infalling time calculations presented here assume a fixed FDM particle mass of $m_{22} = 3$, as in \cite{Hui17}. We have also investigated the impact of adopting $m_{22} = 20$ on our results.

As shown in Table~\ref{Table1}, the differences between the two approaches are substantial. For CDM, our integration consistently yields shorter infalling times than those reported by \cite{Hui17}. For example, for GC1 (GC3) at an initial radius of 1.85 (0.5) kpc, we find an infalling time of 25 (0.24) Gyr, compared to 112 (0.62) Gyr in \cite{Hui17} (see Table~\ref{Table1}). Indeed, it appears that the farther the GC is from the galactic center, the larger the discrepancy between our results and previous estimates becomes. This is because our method accounts for the increasing efficiency of DF as the GC approaches the denser central regions of the NFW halo, where the local density $\rho(r)$ increases significantly. The contrast is even more striking in the FDM case. For the same GC3, they estimated an infalling time of 2.2 Gyr, while our full integration yields 19 Gyr. The infalling times in FDM are systematically longer than in CDM, and significantly longer than those obtained using the constant-coefficient FDM model (see Table~\ref{Table1}).

\begin{figure}
    \centering
    \includegraphics[width=\linewidth]{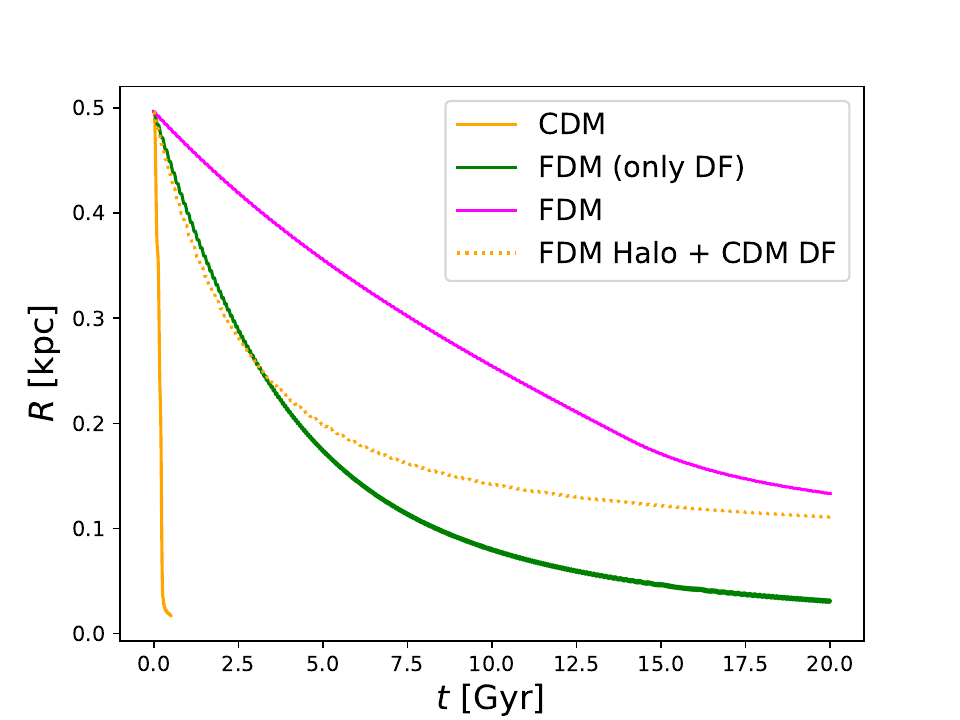}
    \caption{Stalling of Fornax GC3: Orbital radius of GC3 (third row of Table~\ref{Table1}) as a function of time, starting from its currently observed projected radius, assuming a circular orbit as the initial condition. Orange curves show orbital integrations using the classical Chandrasekhar DF formula, while green and magenta curves use the FDM DF model developed in this work. Solid orange and green lines correspond to orbits integrated in a cuspy NFW halo, while dashed orange and solid magenta lines represent orbits in a large-core halo model from \cite{2012MNRAS.426..601C}.}
    \label{GC3}
\end{figure}

More precisely, Figure~\ref{GC3} illustrates how FDM alters the expected fate of GC3. We compared different configurations combining two DM halo profiles and two prescriptions for DF. In the NFW + CDM DF configuration (solid orange line), the cluster rapidly loses energy and spirals into the galactic center in less than 0.5 Gyr. This is expected, as both the steep central density cusp and the classical DF act efficiently to remove energy and angular momentum from the orbit. In contrast, when the same NFW halo is combined with the FDM friction (solid green), the orbital decay is much slower. The cluster gradually falls over several Gyr, but the decay effectively stalls around 0.05–0.1 kpc after 15 Gyr. Contrary to the findings of \citet{BO19} (see their Figure 3), we show that stalling can occur even in the absence of a DM core, solely due to FDM DF (see Figure~\ref{GC3}). We reproduce results similar to theirs only when we artificially fix the friction coefficient $\mathcal{C}_{\rm FDM}$ as a function of time, rather than computing it self-consistently along the orbit. This suggests that their approach, which relies on a Langevin formulation of the Fokker–Planck equation, may underestimate the dynamical heating induced by FDM granules. One possible explanation is that their method assumes circular orbits at each timestep, which prevents them from properly capturing the FDM dynamical heating. For a core FDM-like profile (dashed curves in Figure~\ref{GC3}), the decay slows down further. In the case of a core halo combined with CDM friction (dashed orange line), the infalling remains ongoing but is delayed compared to the cuspy halo case, simply because the lower central density reduces the strength of the friction. Finally, in the most realistic FDM configuration (core halo + FDM DF in solid magenta line in Figure~\ref{GC3}), the orbital decay is extremely inefficient. Even after 20 Gyr, the GC has not reached the center and stabilizes at a radius of about 0.2 kpc. Our results clearly show that FDM DF enhances the stalling effect triggered by the presence of a DM core. The combined effect of a shallow core and FDM-induced heating can effectively cancel the classical infalling. In particular, it shows that Fornax GC3 can survive for more than a Hubble time without requiring fine-tuned initial conditions. This reinforce the idea that FDM is a natural explanation for the survival of Fornax’s GCs, which remains a challenge in the standard CDM framework. Notably, even for higher values of $m_{22}$, such as $m_{22}=20$ (see Table~\ref{Table1}), the timing problem remains alleviated.

In summary, semi-analytical estimates with fixed friction coefficients tend to underestimate the efficiency of DF in CDM and overestimate it in FDM. Overall, our analysis confirms that full orbital integration is essential for accurately estimating GC infalling times, particularly in alternative DM models. In FDM, where heating effects, all the GCs can survive for several Hubble times, even if their orbits lie deep within the halo. This has strong implications for using GCs as probes of DM physics in galactic cores.

\section{Results}

By galactic cannibalism of GCs we refer to the process in which DF drives GCs toward the galactic center, where they are ultimately destroyed by the strong tidal field. This canonical scenario predicts the progressive disappearance of GCs in galaxies over cosmic time. To investigate the impact of FDM DF on GCs over 10~Gyr of evolution, and to identify possible FDM dynamical signatures, we populate the energy-angular momentum space of host halos spanning masses from $10^9$ M$_\odot$ (dwarf galaxies) to $10^{12}$ M$_\odot$ (MW-like galaxies). For each halo, we generate 4500 GCs with initial galactocentric radii uniformly distributed in log space between $0.1$ and $10\,r_s$, where $r_s$ is the scale radius of the host halo. Given the present-day halo mass, $r_s$ is estimated from cosmological $N$-body simulations \citep{2014MNRAS.441.3359D} using \texttt{COLOSSUS} \citep{COLOSSUS}. Velocities are assigned to populate the $E$-$L_z$ diagram, ensuring that all possible bound orbital configurations are represented. All GCs are assigned a mass of $10^6$ M$_\odot$ and a half-mass radius of 10~pc. This choice maximizes the expected effect of DF which is more efficient for compact and massive objects. For reference, $6\times10^6$ M$_\odot$ corresponds to the upper limit of the stellar mass of Galactic GCs about 12~Gyr ago \citep{Baumgardt19}. In both CDM and FDM-DF-only models, the host halo is modeled with an NFW profile, whereas in the FDM case, the DM distribution is replaced with a cored profile as described in Equations~\eqref{galpy1} and ~\eqref{galpy2}. The core radius is computed using the \cite{Schive14} relation at $z=0$ based on FDM cosmological simulations:
\begin{equation}
r_c(M_h, m_{22}) = 1.6 \left( \frac{M_h}{10^9} \right)^{-1/3} m_{22}^{-1},
\label{rcFDM}
\end{equation}
GC orbits are evolved in a fixed potential for 10~Gyr, with integrations performed using fast C integrators \texttt{dopr54\_c} and \texttt{dop853\_c} implemented in the publicly available code \texttt{galpy} \citep{Bovy15}. Regarding the computational performance of our model, orbital integrations over 10~Gyr for 4500 GCs require approximately 0.04~CPU~hours. However, for our FDM-DF-only and FDM models, Figure~\ref{CT} shows how the computational time depends on the FDM particle mass $m_{22}$. The difference in computational time between both models is not due to the use of the FDM core itself, but rather to the fact that the \texttt{TwoPowerSphericalPotential} is not implemented in C.

\begin{figure}
    \centering
    \includegraphics[width=\linewidth]{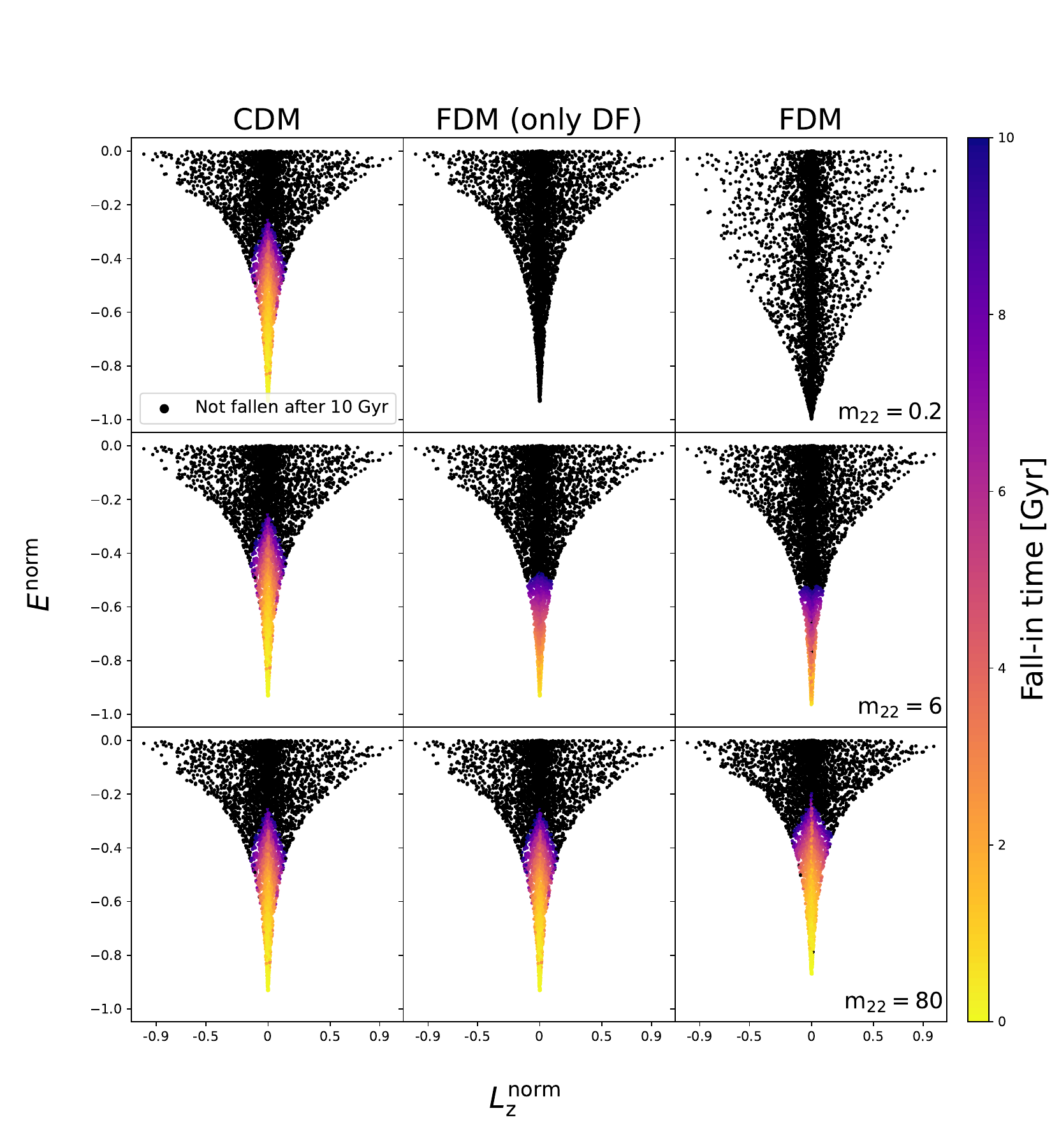}
    \caption{Normalized total initial energy as a function of the normalized initial $z$-component of the angular momentum at $z=0$, color-coded by the infall time, for $10^6$ M$_\odot$ GCs orbiting within $10^9$ M$_\odot$ DM halos. Results are shown for CDM (left panels), FDM with only DF (middle panels), and FDM (right panels) with $m_{22} = 0.2, \, 6, \,$ and $80$. The infall time is defined as the time at which the apocentre of the GC orbit drops below $r_{\rm lim} = 0.157 \, \mathrm{kpc}$ (10\% of the scale radius of a $10^9$ M$_\odot$ NFW halo). A total of 4500 gravitationally bound GCs were randomly initialized in logarithmic radius between $0.1$ and $10 \, r_s$. The energy is normalized to the absolute value of the minimum of the gravitational potential, $|E_{\rm min}|$, and the angular momentum to the absolute value of the maximum angular momentum in our GC sample. The first column corresponds to the CDM case with Chandrasekhar DF.}
    \label{fig4}
\end{figure}

\begin{figure*}
    \centering
    \includegraphics[width=0.8\linewidth]{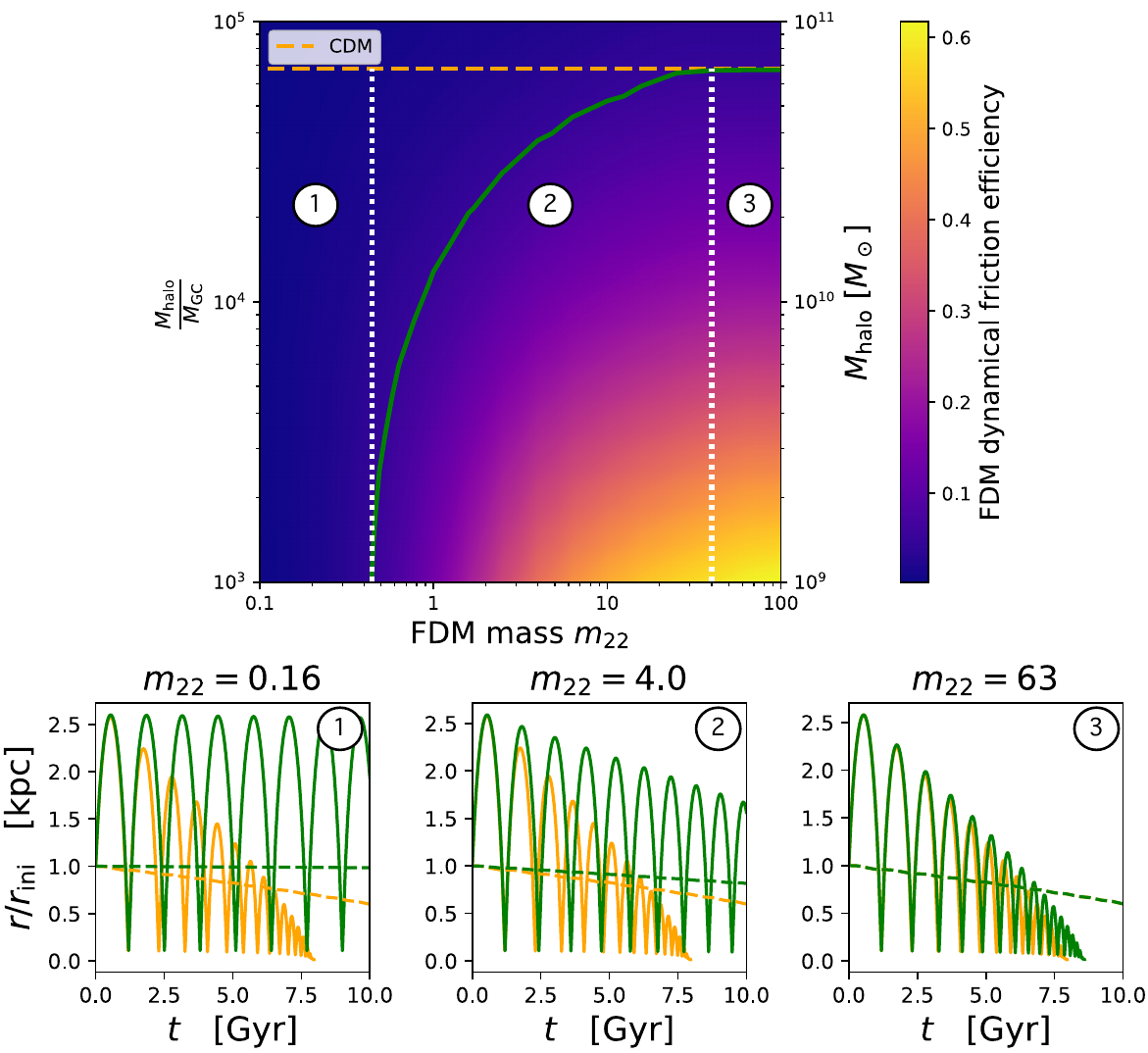}
    \caption{Top panel: DF efficiency map for FDM-DF-only model with halo-to-GC mass ratios between $10^3$ and $10^5$, and FDM particle masses $m_{22}$ ranging from 0.1 to 100. The secondary axis shows the corresponding halo mass assuming $M_{\mathrm{GC}} = 10^{6}$ M$_\odot$. The map is computed on a $16 \times 16$ grid, where each point corresponds to a NFW halo of a given mass and a DF force derived for a specific FDM particle mass. For each combination of halo-to-GC mass ratio and FDM particle mass, the DF efficiency is evaluated using 1000 orbits in order to limit computational cost. There were uniformly distributed in logarithmic space between $0.1$ and $10\,r_s$. The orange dashed and green solid lines indicate the 5\% efficiency frontier for CDM and FDM-DF-only models, respectively. Bottom panels: Orbital radius normalized to its initial value as a function of time, for GCs integrated over 10 Gyr in a $10^9$ M$_\odot$ NFW halo with classical DF (orange curves) and FDM DF (green curves) for $m_{22} = 0.16,\, 4,$ and $63$. For each case, both radial and circular orbits are shown. Dashed white lines separate the three regimes of FDM DF: (1) absence of friction inducing orbital stalling, (2) reduced friction, and (3) the classical regime.}
    \label{fig5}
\end{figure*}

\begin{figure}
    \centering
    \includegraphics[width=\linewidth]{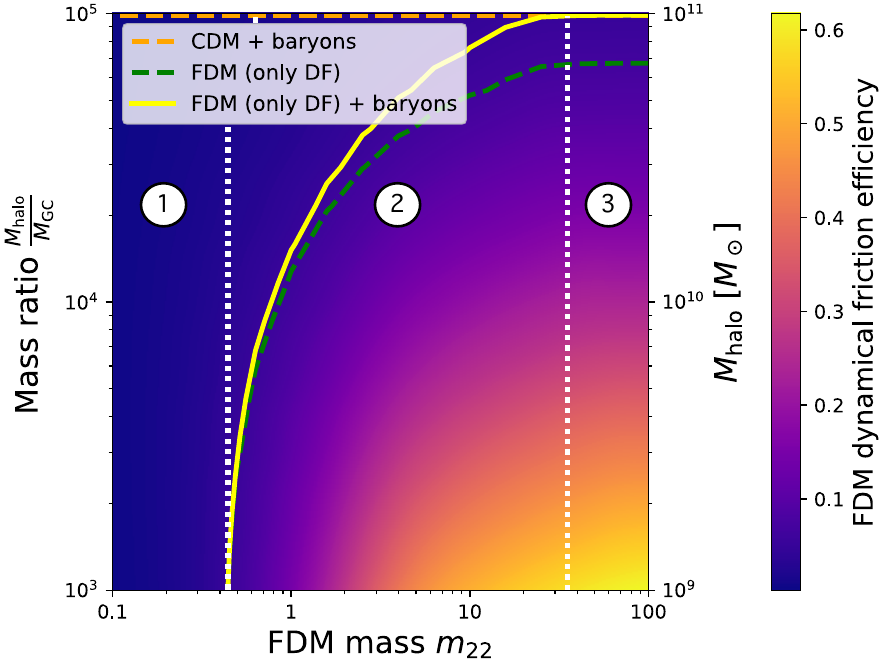}
    \caption{Impact of baryons: DF efficiency map for FDM-DF-only + baryons model with halo-to-GC mass ratios between $10^3$ and $10^5$, and FDM particle masses $m_{22}$ ranging from 0.1 to 100. The secondary axis shows the corresponding halo mass assuming $M_{\mathrm{GC}} = 10^{6}$ M$_\odot$. Dashed white lines separate the three regimes of FDM DF: (1) absence of friction, (2) reduced friction, and (3) the classical regime.}
    \label{fig6}
\end{figure}

\subsection{FDM dynamical friction efficiency}

Figure~\ref{fig4} shows the orbital evolution of our GCs in dwarf galaxy halos of mass $10^9$ M$_\odot$, under different DM scenarios and FDM particle masses. The panels display the normalized energy and angular momentum as a function of the infall time toward the galactic center. The energy is normalized to the absolute value of the minimum of the gravitational potential, $|E_{\rm min}|$, and the angular momentum to the absolute value of the maximum angular momentum in our GC sample. Each point corresponds to a GC, color-coded by the time it takes for the apocenter of its orbit to fall below \(r_{\text{lim}} = 0.157\,\mathrm{kpc}\) (10\% of the scale radius of a $10^9$ M$_\odot$ NFW halo). Yellow points indicate rapid infall (less than 2 Gyr), dark blue points indicate slow infall (more than 8 Gyr), while black points correspond to GCs that do not fall in within 10 Gyr. The first column of Figure~\ref{fig4} corresponds to the CDM case with Chandrasekhar DF. Here, friction is efficient and drives the rapid infall of GCs on tightly bound, low-angular-momentum orbits, while those with higher energies and angular momenta remain on wide orbits. The second column of Figure~\ref{fig4} shows the case of FDM with FDM-specific DF but assuming a standard NFW potential. In this case, the friction is much weaker, especially for \(m_{22} = 0.2\), where no GC falls in at all. For \(m_{22} = 6.0\), unlike CDM, only the GCs with energies lower than $-0.5\;|E_{\rm min}|$ reach the galaxy center within 10 Gyr (see Figure~\ref{fig4}). Finally, for \(m_{22} = 80\), the results converge back to the CDM case. The third column illustrates the FDM scenario with a central core modeled with Equations~\eqref{galpy1} and ~\eqref{galpy2}. The flattened density profile in the center further reduces the efficiency of DF, such that even low-energy GCs experience longer infall times (see FDM $m_{22} = 6.0$ in Figure~\ref{fig4}). As expected, smaller values of \(m_{22}\) generate larger cores, resulting in even weaker friction and an almost complete absence of infall; conversely, for high \(m_{22}\), the profile approaches that of a CDM halo and the behavior becomes indistinguishable from the classical case.  

For $m_{22} = 6$, the CDM column in Figure~\ref{fig4} clearly shows a color ridge pointing toward high energies: the infall time increases with angular momentum $L_z$, as expected from standard DF, because on nearly circular orbits (high $L_z$) GCs remain at roughly constant radius where the density is lower and the relative velocity nearly constant, so energy loss is slow, whereas on more radial orbits (low $L_z$) GC plunge through the dense central regions where DF is strongest, leading to rapid orbital decay. By contrast, in the FDM-DF-only case ($m_{22} = 6$) with an NFW potential and even more so in the FDM core case, the color distribution is nearly inverted, producing a ridge oriented toward negative energies (see Figure~\ref{fig4}). As shown for both gas dynamical friction \citep{2023DDA....5420005G,eytan2025harmonicdecompositionapproachdynamicalfriction} and for FDM gravitational atoms \citep{Tomaselli_2023}, dynamical friction in a steep density cusp tends to circularize orbits, while in a shallow or nearly uniform density profile it instead drives orbits toward higher eccentricity. In a cuspy profile, the strong density contrast between pericentre and apocentre causes energy loss to be dominated by passages through the central regions, leading to a preferential damping of radial motions and hence circularization. In contrast, in a cored potential the density varies weakly along the orbit. In the latter, the GC spends more time near apocentre than near pericentre thus energy loss accumulated at apocentre dominates, enhancing orbital eccentricity. The color inversion in Figure~\ref{fig4} suggests that at a given energy, eccentric orbits are more strongly heated by FDM fluctuations, since the GC passes close to the center where heating efficiently counteracts DF. The resulting stalling is therefore more effective than for circular orbits. This behavior directly echoes our Figure~\ref{figure2}, which clearly shows that for a given $m_{22}$ a GC starting from the same apocentre takes longer to reach the center on a circular orbit than on an eccentric one. The presence of an FDM core therefore naturally amplifies the inversion of the color distribution, reinforcing the tendency for eccentric orbits to stall more efficiently than circular ones.
 
We explored these behaviors for halo-to-GC mass ratios between $10^3$ and $10^5$, and for FDM particle masses $m_{22}$ ranging from 0.1 to 100. Rather than computing the infall time, which would require an arbitrary choice of radius in the absence of mass loss, we chose instead to calculate the efficiency of DF, defined as the mean relative difference between the last apocentre without DF and with the FDM DF. The advantage of this metric is that it provides a robust measure of the impact of DF on orbital evolution, independent of the GC initial conditions and the DM distribution. We set a threshold at 5$\%$. Below this value, DF is considered no longer efficient, and this constitutes our efficiency limit for all models.  

Figure~\ref{fig5} illustrates the efficiency of FDM DF in an NFW halo as a function of the particle mass \(m_{22}\) and the mass ratio \(M_{\rm halo}/M_{\rm GC}\), along with the typical orbital evolution for three representative values of \(m_{22}\). For comparison, we also compute the efficiency boundary for classical DF using Chandrasekhar’s formula, which depends only on the mass ratio. We find that above a mass ratio of $7\times 10^4$, friction becomes negligible (shown as an orange dashed line in Figure~\ref{fig5}). The three lower panels in Figure~\ref{fig5} show the decay of the normalized orbital radius (relative to its initial value) as a function of time for a GC in an NFW halo, for three representative values of \(m_{22}\) corresponding to the identified regimes. Green curves correspond to the FDM-DF-only model, orange to CDM. For each model, both radial and circular orbits are shown. Similar to CDM, our results confirm that the galactic regime of interest for significant FDM signatures corresponds to dwarf galaxies with $M_{\rm halo} = 10^9-10^{10}$, where the mass ratio between GCs and the host galaxy is lowest. In this regime, the free parameter of FDM, $m_{22}$, induces clear differences, which lead us to identify three distinct zones. (1) zone 1, to the left of the FDM DF efficiency boundary (green line in Figure~\ref{fig5}), highlights a regime where DF is negligible regardless of the relative mass, meaning that on average GCs lose virtually no orbital energy. Since significant DF effects are found below a mass ratio of $10^4$, we restrict this zone to $m_{22} \leq 0.44$ (white dashed line in Figure~\ref{fig5}). For \(m_{22} = 0.16\), FDM friction is so weak that the orbit retains nearly constant radial amplitude over 10 Gyr, while the CDM case shows a clear decay. This illustrates the DF stalling mechanism that naturally arises in our FDM-DF-only model, even in the absence of DM core. (2) zone 2, below the FDM DF efficiency boundary, corresponds to a regime where friction remains effective but reduced compared to CDM. The bottom middle panel with \(m_{22} = 4.0\) confirms this with orbital decay in FDM, though less pronounced than in CDM. (3) Finally, in zone 3, for $m_{22} > 40$, friction becomes increasingly similar to classical DF, with total convergence toward $m_{22}=80$. For \(m_{22} = 63\), the orbital decay is nearly identical to CDM, reflecting the recovery of the classical regime where quantum effects are negligible.

In summary, Figure~\ref{fig5} shows that in the dwarf regime, FDM DF is highly inefficient for low \(m_{22}\) inducing orbital stalling, reduced for intermediate values (\(0.45 \le m_{22} \le 40\)) where DF dominates over FDM granular heating, and converges back to the CDM behavior at high \(m_{22}\). The existence of such regimes naturally leads to the emergence of a stalling mechanism: for sufficiently low $m_{22}$, GCs cease to lose orbital energy and remain on wide orbits, escaping the fate of infall. This mechanism therefore directly challenges the expected galactic cannibalism of GCs, and provides a pathway for their long-term survival in dwarf galaxies.

\subsection{Impact of baryons}

Here, we aim to estimate how our results are affected by the presence of baryons, which increase the central density of the galaxy and thus enhance DF according to Equation~\eqref{eq:ChandraDF}. Given the uncertainty in the stellar distribution within FDM halos, we adopt the assumption that it follows the CDM model at $z=0$. This is why, our goal is to provide a qualitative description of the impact of baryons on the boundaries of FDM DF efficiency. The potential of the stellar component is characterized by its mass and scale radius. We therefore assume the stellar-to-halo mass relation from abundance matching \citep{2010ApJ...717..379B} to estimate the baryonic mass of our galaxies using \texttt{Halotools}\footnote{\url{https://halotools.readthedocs.io/en/latest/index.html}} 
(v0.9, \citealt{2017AJ....154..190H}). The effective radius is then calculated following Equation~32 of \cite{2003MNRAS.343..978S}. The half-mass radius of the stellar distribution is assumed to be $\rhm = 0.75 \, r_{\rm eff}$ \citep{1990ApJ...356..359H,2013ApJ...763...73S}. In the case of classical DF, the efficiency boundary is shifted from a mass ratio of $7 \times 10^4$ to $10^5$ in the presence of baryons (see Figures~\ref{fig5} and \ref{fig6}).  

Figure~\ref{fig6} shows the impact of the baryonic component on FDM DF. As in Figure~\ref{fig5}, three regimes can be distinguished: inefficiency for low masses ($m_{22}<0.44$), reduced friction ($0.44 \lesssim m_{22} \lesssim 30$), and convergence toward the CDM regime for $m_{22}>30$. Overall, the addition of baryons broadens the range of mass ratios over which DF is efficient, since the extra central density enhances orbital energy losses. In particular, convergence toward the CDM case is now reached at a mass ratio of $10^5$ for large $m_{22}$. Moreover, the baryonic component enlarges the CDM-like regime (zone~3) while reducing the extent of zone~2, where DF is active but reduced.

\begin{figure}
    \centering
    \includegraphics[width=\linewidth]{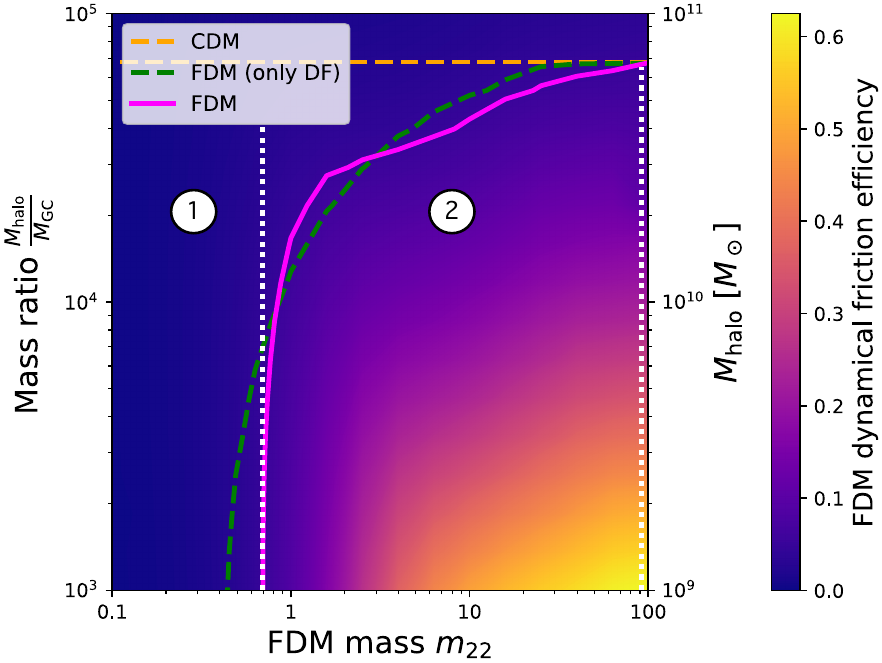}
    \caption{Impact of DM core: Same as Figure but for FDM model with halo-to-GC mass ratios between $10^3$ and $10^5$, and FDM particle masses $m_{22}$ ranging from 0.1 to 100. The secondary axis shows the corresponding halo mass assuming $M_{\mathrm{GC}} = 10^{6}$ M$_\odot$. Dashed white lines separate the three regimes of FDM DF: (1) absence of friction and (2) reduced friction.}
    \label{fig7}
\end{figure}

\begin{figure*}
    \centering
    \includegraphics[width=\linewidth]{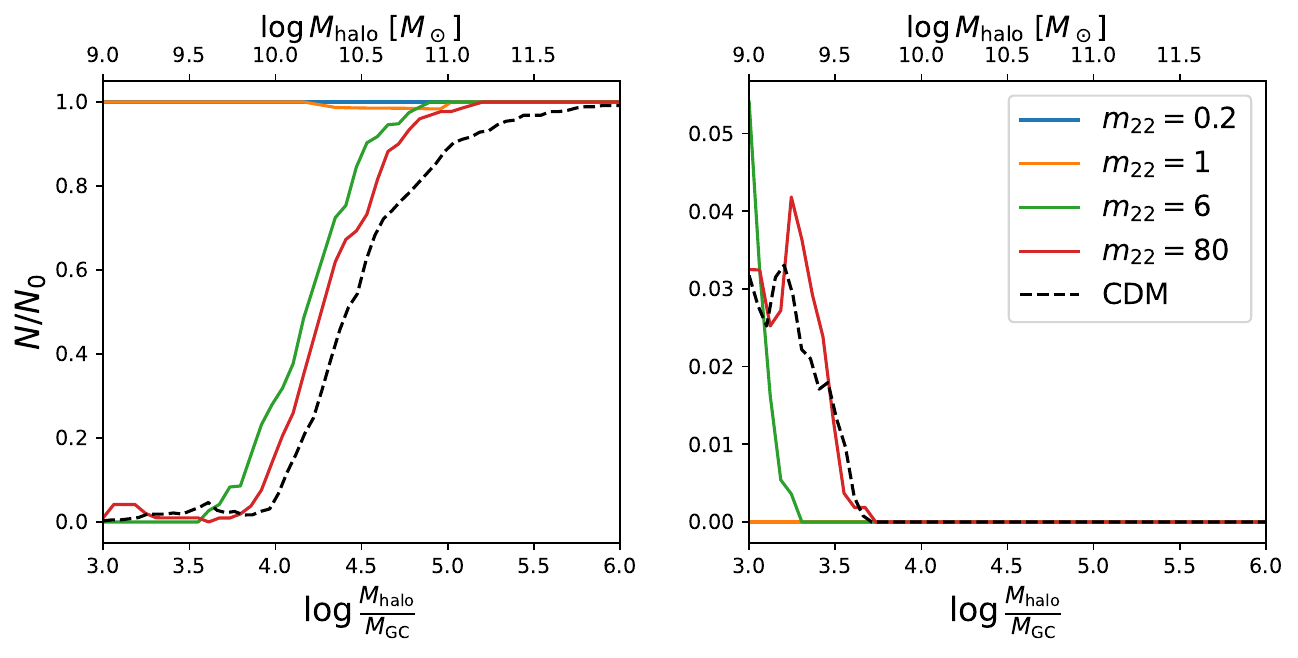}
    \caption{Ratio of surviving to initial GCs within the central region of galaxies delimited by the scale radius $r_s$ of halos as a function of the mass ratio $M_{\rm halo}/M_{\rm GC}$ and for different FDM particle masses $m_{22}$. The survival criterion corresponds to a final apocenter larger than $0.1\,r_s$. The left panel shows an in-situ population ($r_{\rm apo}^{\rm init} < r_s$) initially composed of about $N_0=300$ GCs, while the right panel corresponds to an ex-situ population ($3r_s < r_{\rm apo}^{\rm init} < 10r_s$) initially composed of about $N_0=2100$ GCs.}
    \label{fig8}
\end{figure*}

\subsection{Impact of FDM core profile}

Figure~\ref{fig7} shows the DF efficiency map in our FDM model, this time including the FDM DF and the density profile modified by the presence of a solitonic core. Compared to the FDM-DF-only model, the introduction of the FDM core profile significantly alters the dynamics in the central regions. In Figure~\ref{fig7}, the oscillation of the FDM boundary (purple line) around the FDM-DF-only case (dashed green line) arises from the shape of the FDM profile, which can be more or less dense than NFW profile depending on the core radius and central density, themselves determined by $m_{22}$ and $M_{\rm halo}$. Only two regimes can be clearly distinguished as a function of $m_{22}$. The inclusion of the central core strengthens the DF inefficiency for dwarf galaxies up to $m_{22} = 0.7$, and nearly suppresses the convergence toward the CDM regime (zone~3) over the considered FDM mass range. Instead, it substantially enlarges the regime in which DF is reduced by quantum heating. This implies that in a realistic FDM scenario, accounting for both FDM DF and the solitonic core structure, the expected outcome is a population of GCs much more resilient to galactic cannibalism.

\section{Demographics of GC systems}

This stalling mechanism, arising both from quantum heating counteracting DF and from the presence of a DM core, has direct implications for the galactic cannibalism of GCs: by preventing their orbital decay and subsequent tidal destruction near the center, it ensures the long-term survival of a significant fraction of GCs. In other words, the properties of FDM may shield GCs from the canonical cannibalism scenario expected in CDM halos, thereby reshaping the predicted demographics of GC systems. To explore the impact of FDM on GC system demographics, we artificially construct an in-situ and an ex-situ GC population within our galaxies, which should be regarded as a limiting case. At the initial time, in-situ GCs are defined as having apocenters $r_{\rm apo}^{\rm init} < r_s$, while ex-situ GCs have $3r_s < r_{\rm apo}^{\rm init} < 10r_s$. The in-situ population forms preferentially in the central regions of galaxies, where the gas density—and thus the likelihood of massive cluster formation—is highest, whereas the ex-situ population consists of GC accreted through mergers, naturally residing at larger distances from the center. These GCs are selected from our initial sample of 4500 clusters. For our orbital analysis, we define a GC to have sunk to the center once its apocenter falls below 10\% of the scale radius of its host halo. This criterion naturally depends on the total halo mass: for example, in the MW , where the DM scale radius is about 16 kpc \citep{Bovy15}, no GC is found within 700 pc. Moreover, in the Fornax dwarf galaxy, no GC is found within 0.14 kpc, which corresponds to 10\% of the scale radius of a $8 \times 10^8$ M$_\odot$ DM halo. This criterion thus allows us to determine whether a GC is destroyed, given that our model does not include GC mass loss.


\subsection{Absence of mixing of in-situ and ex-situ populations}

Figure~\ref{fig8} illustrates the ratio of surviving in-situ to initial in-situ GCs (left panel), and the ratio of ex-situ surviving in-situ to initial ex-situ GCs (right panel) within the central region of our galaxies (inside the scale radius $r_s$ of the halos) as a function of their mass ratio $M_{\rm halo}/M_{\rm GC}$, for different values of the FDM particle mass $m_{22}$. After 10~Gyr of evolution, both panels of Figure~\ref{fig8} again reveal that the mass regime of interest is that of dwarf galaxies, i.e. between $10^9$ and $10^{10}$ M$_\odot$, as this is where the strongest discrepancies between FDM and CDM appear. This arises because, in this regime, DF is maximized in both CDM and FDM, but FDM heating can counteract it with an efficiency that depends on $m_{22}$. In the in-situ case (left panel), GCs are subjected to particularly strong CDM DF as the mass ratio decreases, leading to an almost complete disappearance of the in-situ population once $M_{\rm halo}/M_{\rm GC}$ drops below $10^{4}$ (see black dashed line in Figure~\ref{fig8}). This behaviour is altered in FDM and strongly depends on $m_{22}$: for $m_{22}<4$, DF is inefficient and most clusters survive; for $4 \lesssim m_{22} \lesssim 30$, friction is enhanced and a large fraction of clusters sink; finally, for $m_{22}>30$, the regime tends to recover the CDM behaviour, with a near extinction of the in-situ population in dwarfs. In the ex-situ case (right panel), the majority of GCs initially located beyond $3r_s$ do not sink after 10~Gyr and remain outside the central region of our galaxies (delimited by $r_s$) for mass ratios above 4. However, in the dwarf regime ($\log(M_{\rm halo}/M_{\rm GC})=3-4$), about 3\% of GCs (75 GCs) penetrate the central region through DF and survive there in CDM. Although the fraction of ex-situ clusters populating the centre appears small, they represent a non-negligible 20\% compared to the in-situ clusters initially located at the centre (300 clusters at $t=0$). For $m_{22}<6$, none at all, are found in the central region due to friction being halted by quantum heating. A special case arises for $m_{22}\sim6$, where friction is efficient enough to bring clusters below $r_s$, but once at the centre quantum effects become strong enough to induce orbital stalling. This results in a GC ratio (5\%) higher than in CDM (3\%).

\begin{figure}
    \centering
    \includegraphics[width=\linewidth]{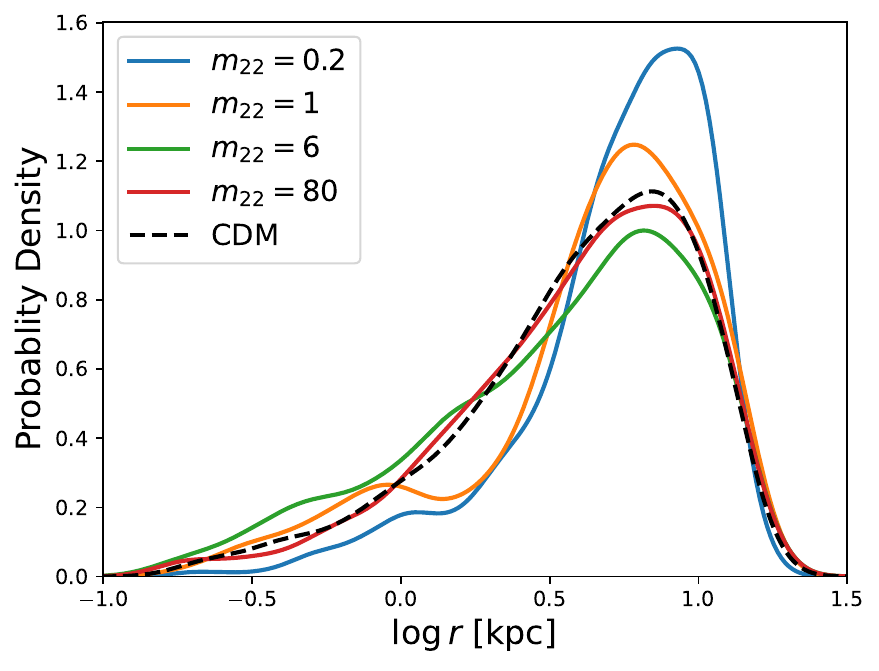}
    \caption{Limited mixing of the two GC populations: Radial probability density distributions of GCs for the total population (in-situ + ex-situ) in a dwarf galaxy with a mass ratio of $10^3$. For low values of $m_{22}$, the total distribution exhibits a clear FDM signature with a bimodal shape, characterized by two distinct peaks and a gap between them. This feature reflects the limited mixing between the in-situ and ex-situ populations induced by the FDM DF and the DM core. As $m_{22}$ increases, this bimodality progressively disappears and the distribution converges towards the CDM case, indicating enhanced mixing of the two populations.}
    \label{fig9}
\end{figure}

For dwarf galaxies, our results therefore show that FDM favours a dominance of in-situ clusters in the central region, whereas CDM favours a dominance of ex-situ clusters. The survival of in-situ GCs and the near absence of ex-situ clusters in the central regions of dwarfs under FDM, compared to CDM, highlights a crucial point. An unambiguous identification of in-situ clusters in dwarf galaxies would provide strong constraints on the free parameter $m_{22}$, whose value remains highly debated between galactic and cosmological scales. Indeed, galactic studies, such as the analysis of the thickening of cold stellar streams \citep{2018arXiv180800464A}, suggest $m_{22} > 1.5$ and investigations of the MW disk indicate a boson mass around $m_{22} \sim 0.5 - 0.7$ and $m_{22} > 1.3 $ by \cite{2023MNRAS.518.4045C} and \cite{2024arXiv241213275H}, respectively. These findings are in tension with the more stringent limits derived from the Lyman-$\alpha$ forest, which constrain the boson mass to $m_{22} = 7 - 20$ \citep{2017PhRvL.119c1302I,2017PhRvD..96l3514K,2019MNRAS.482.3227N,2017MNRAS.471.4606A,2021PhRvL.126g1302R}. According to Figure~\ref{fig8}, galactic-scale constraints support the idea that GCs may act as an effective probe to discriminate between CDM and FDM, whereas large-scale constraints place us in a regime where GCs show no significant deviations from CDM.

Our results also demonstrate that the FDM scenario can naturally reconcile the Fornax problem and its six GCs \citep{Mackey03} (Fornax GC1–5) and \cite{Pace21} (Fornax GC6). In the FDM framework, these GC could be in-situ and survive over a Hubble time due to reduced or even suppressed DF. In contrast, in CDM, the dynamical history of the GCs is more complex. Most of Fornax's GCs would need to be ex-situ, which would require a merger in Fornax's history with a lower-mass galaxy capable of delivering these GCs. Indeed, \cite{Deason14} found that 10\% of satellite dwarf galaxies with $M_\star > 10^6$ M$_\odot$ within the host's virial radius experienced a major merger with a stellar mass ratio larger than 0.1 since $z=1$. Although the probability that this occurred specifically for Fornax is low, it is notable that Fornax is the only one of the eleven classical MW satellites to host GCs, pointing to a potential past merger. 

Such a merger would have served two purposes. First, it could have brought in ex-situ clusters that could survive in a galaxy where tidal cannibalism is intense. Indeed, more than 50\% of dwarf galaxies with $M_\star < 2 \times 10^7$ M$_\odot$ do not host GCs, and even if they do, it is rarely more than three for $10^7$ M$_\odot$ \citep{Chen23}. Second, the merger could have triggered orbital expansion of in-situ clusters, allowing them to survive longer \citep{Leung20}. However, this mechanism only works for three clusters (GC1, GC2, GC3) among the five Fornax GCs studied by \cite{Leung20}. The evidence for a past merger therefore seems insufficient and may even bias our understanding, as it does not guarantee that the accreted galaxy contributed any GCs to Fornax. Again, we emphasize that determining the origin of the clusters could be a decisive metric to address this timing problem.


\subsection{Bimodality in the GC radial distribution}

The FDM signatures in the GC dynamics shown in Figure~\ref{fig8} for low $m_{22}$ values leads to very limited mixing between the in-situ and ex-situ populations in the central region, as would be expected in CDM. This contrasting behaviour of in-situ and ex-situ populations in CDM and FDM scenarios demonstrates a potential signature in the radial distribution of clusters in an FDM universe within the dwarf galaxy regime, as illustrated in Figure~\ref{fig9}. This figure shows the total GC population after 10~Gyr of evolution, artificially created in a dwarf galaxy with a mass ratio of $10^3$. Unlike in CDM, which shows an indiscernible mixture of populations, the distinction between in-situ and ex-situ clusters in FDM becomes increasingly clear as $m_{22}$ decreases. In-situ GCs occupy the central regions of the galaxy, whereas ex-situ clusters remain at larger radii. This absence of galactic cannibalism results, for the total population, in a bimodal radial distribution when considering low FDM particle masses ($m_{22} < 6$). Conversely, for higher values of $m_{22}$, the distribution becomes progressively smoother, gradually reproducing the behaviour expected in the CDM scenario, with a stronger overlap of the two populations and a disappearance of the bimodal feature.

Thus, the presence or absence of this bimodal signature in the total GC population could serve as a sensitive indicator of the nature of DM. The ongoing Euclid mission will be able to provide the projected spatial distribution of GCs across its unprecedented coverage of galaxies, spanning stellar masses from $10^{9}$ M$_\odot$ (dwarfs) to $10^{12}$ M$_\odot$ (MW-like systems) in the nearby Universe ($< 100$~Mpc), by detecting half a million extragalactic GCs \citep{2021sf2a.conf..447L,Voggel25}. In the dwarf regime, the Euclid early release observations of the Perseus galaxy cluster have already enabled the successful detection of 11370 candidate GCs \citep{2025arXiv250316367S}, associated with more than 1000 dwarf galaxies with stellar masses between $10^6$ and $10^{10}$ M$_\odot$ \citep{2025A&A...697A..12M}. More statistics across the full stellar mass range of dwarfs with the upcoming Euclid DR1 in autumn 2026 could be sufficient to investigate this signature. However, this signature still needs to be confirmed by taking into account hierarchical assembly with mass loss including contributions from stellar evolution, two-body relaxation, and tidal shocks but also and more realistic initial conditions for GCs, as in \cite{Boldrini25}.


\section{Conclusions}

In this work, we have implemented the formalism of FDM DF in the \texttt{galpy} framework, enabling for the first time full orbital integrations of GCs under this alternative DM scenario. Our results confirms that the wave-like nature of FDM fundamentally alters the efficiency of DF compared to the standard CDM paradigm. The efficiency of FDM DF depends sensitively on the boson mass $m_{22}$. We identify three regimes: (i) at low $m_{22}$, DF is suppressed and GCs remain at large radii; (ii) at intermediate $m_{22}$, DF is partially effective but reduced compared to CDM; (iii) at high $m_{22}$, the classical CDM behaviour is recovered. These regimes open a new pathway to constrain $m_{22}$ on galactic scales, in complement to cosmological probes such as the Lyman-$\alpha$ forest.

We show that stochastic heating by FDM density granules, combined with the presence of solitonic cores, induces a stalling mechanism that prevents the long-term orbital decay of GCs, most notably in dwarf galaxies where classical DF is expected to be most efficient. This process halts the canonical galactic cannibalism expected in CDM, ensuring the survival of in-situ GCs over a Hubble time and strongly limiting the mixing with ex-situ clusters. As a consequence, GC systems in FDM dwarf halos naturally develop a bimodal radial distribution, with central in-situ clusters coexisting with outer ex-situ ones. Future work should focus on confirming this bimodality signature in dwarfs by taking into account hierarchical assembly with mass loss, as in \citet{Boldrini25}. Importantly, these predictions could provide a robust explanation for the long-standing survival of the Fornax GCs. From an observational perspective, our results highlight GCs as powerful dynamical tracers of FDM physics. With its unprecedented coverage, the Euclid mission will soon deliver the first statistical census of extragalactic GCs, offering a unique opportunity to test these predictions and place new constraints on the nature of DM. 

Crucially, FDM stands out from other alternative DM models such as warm DM \citep{2001ApJ...556...93B} or self-interacting DM \citep{2000PhRvL..84.3760S}. While these scenarios can also produce DM cores in dwarf galaxies, the formulation of DF remains essentially classical and is well described by the Chandrasekhar prescription. In SIDM, for instance, the Knudsen number (${\rm Kn} \gg 1$ for $\sigma/m = 0.2$–$100$ cm$^2$ g$^{-1}$) indicates that gravity dominates over scattering, keeping DF in the classical regime \citep{2024A&A...690A.299F}. However, the presence of a core may reduce the efficiency of DF and lead to core stalling, but this effect arises purely from the DM density structure such as in SIDM \cite{2025arXiv251114912V}. In contrast, FDM uniquely modifies the very mechanism of DF through quantum granule-induced heating, offering a distinctive dynamical signature absent in other models. Identifying such observational signatures is therefore a crucial step forward in the long-term effort to distinguish between DM scenarios and possibly rule some of them out. We also stress that baryons play a non-negligible role in shaping the distribution of DM through both gravitational and hydrodynamical processes. Stellar feedback, for example, can also generate cores in dwarf galaxies, potentially mimicking some DM-induced effects. Disentangling these contributions is essential for a robust interpretation of DF signatures.

\section{Data Availability}

The data underlying this article is available through reasonable request to the author. Codes and data will be available at the following URL: \href{https://github.com/Adrian998obs/FuzzyDarkMatterDynamicalFriction}{github.com/Adrian998obs/FuzzyDarkMatterDynamicalFriction}. The documentation of FDM DF is already available on \texttt{galpy} website at : \href{https://docs.galpy.org/en/latest/reference/potentialfdmdynfric.html}{potentialfdmdynfric}.

\begin{acknowledgements} 

PB acknowledges funding from the CNES post-doctoral fellowship program. This work was supported by CNES, focused on the Gaia mission. PB and PDM are grateful to the "Action Thématique de Cosmologie et Galaxies (ATCG), Programme National ASTRO of the INSU (Institut National des Sciences de l'Univers) for supporting this research, in the framework of the project "Coevolution of globular clusters and dwarf galaxies, in the context of hierarchical galaxy formation: from the Milky Way to the nearby Universe" (PI: A. Lançon). PB thanks members of the Euclid working group, "Extragalactic globular clusters", for contributing with useful comments in the early stages of this project. JB acknowledges financial support from NSERC (funding reference number RGPIN-2020-04712). AS gratefully acknowledges the SUTS Master’s program at the Observatoire de Paris for providing him with the opportunity to carry out this internship. 

\end{acknowledgements}

\bibliography{src}

\appendix

\section{Appendix}

\begin{figure}[t]
    \centering
    \includegraphics[width=1\linewidth]{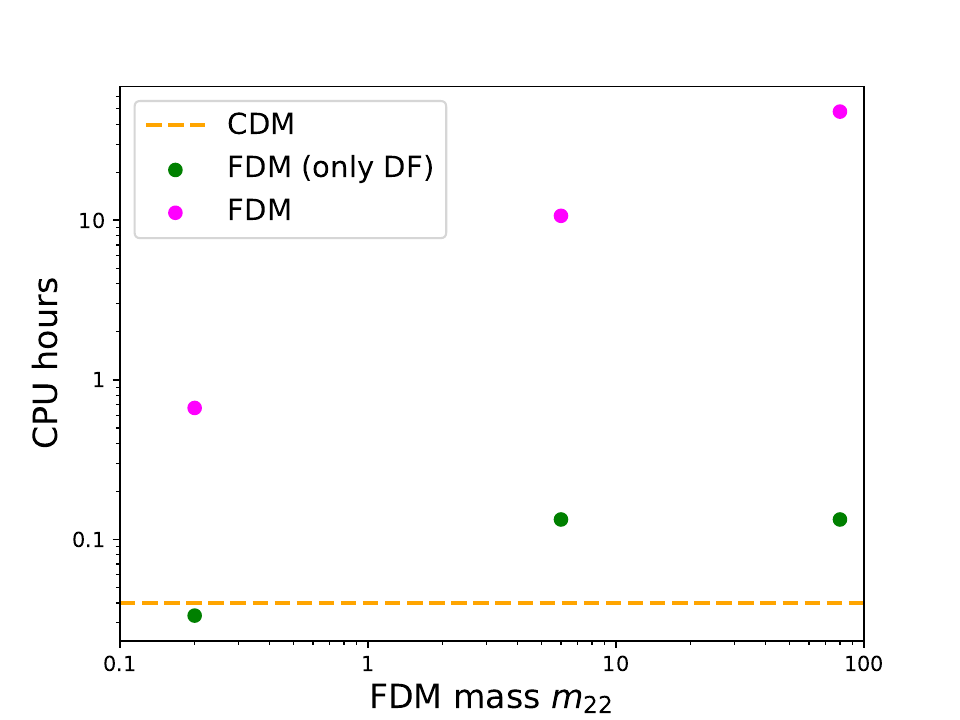}
    \caption{Computational time in CPU hours as function of the FDM particle mass for FDM-DF-only and FDM models. Orange dashed lines represents the computational time for CDM.}
    \label{CT}
\end{figure}

\subsection{Implementation in \texttt{galpy}}
\label{A1}

In the public code \texttt{galpy}, designed for orbit integration in both 
\texttt{Python} and C and widely used in galactic dynamics, DF and galactic potentials are implemented as separate potential classes that can be modified. Classical DF is implemented as a \texttt{Force} instance (\texttt{ChandrasekharDynamicalFrictionForce}). In this work, we have developed a new subclass, \texttt{FDMDynamicalFrictionForce}, based on the formulation of \cite{Lancaster20}, implemented in both \texttt{Python} and C and derived from the classical case. This new subclass introduces the explicit dependence on the parameters $kr$ and $M_\sigma$ for a point-mass object such as a GC. At each timestep, given the values of these parameters computed using Equations~\ref{k} and~\ref{Msigma}, the DF coefficient is evaluated as follows:  

\begin{enumerate}
    \item For $kr < M_\sigma/2$, DF is computed within the FDM zero-velocity dispersion regime (Equation~\ref{0-disp}), shown in cyan in Figure~\ref{fig:Crel}.  
    \item For $kr > 2 M_\sigma$, the coefficient is taken from the FDM dispersion regime (Figure~\ref{disp}), represented in blue in Figure~\ref{fig:Crel}.  
    \item For intermediate values of $kr$, where the behavior of FDM DF is more uncertain, we assume a linear interpolation between the zero-dispersion and dispersion regimes, evaluated respectively at $kr=M_\sigma/2$ and $kr=2M_\sigma$.  
    \item Finally, the classical DF coefficient $\Ccdm$ is computed, and we 
    ensure that the FDM coefficient never exceeds the classical value (see red curve in Figure~\ref{fig:Crel}).  
\end{enumerate}

We chose to compare the FDM DF coefficient $\mathcal{C}_{\rm FDM}$ directly with the classical coefficient $\mathcal{C}_{\rm CDM}$ in order to handle the transition between the two regimes. This approach guarantees that the appropriate regime is applied, classical ($M_Q \ll 1$) or FDM ($M_Q \gg 1$). Figure~\ref{fig:Crel} illustrates the different DF regimes as a function of $kr$, together with the resulting coefficient. Since DF is evaluated at each timestep, an orbit can pass through different regimes during the integration. This represents the main difference with the work of \cite{Hui17}, where only the initial value of $\mathcal{C}_{\rm FDM}$ in the zero-velocity dispersion regime was considered. This feature leads to significant differences in orbital evolution. Further examples and technical details of the class are provided in the \texttt{galpy} documentation\footnote{\url{https://docs.galpy.org/en/latest/reference/potentialfdmdynfric.html}}.

\subsection{Validity tests}
\label{A2}

In order to validate the new \texttt{FDMDynamicalFrictionForce} class, we verified both the consistency of the computations and the behavior of the DF force through a series of numerical tests. First, we confirm that in the classical regime (high $\mFDM$), FDM DF converges to the \texttt{ChandrasekharDynamicalFrictionForce}, with a relative error in the final positions below 0.1\%, after 2 Gyr. Second, we show that in the central limit ($kr \ll 1$), FDM DF agrees with analytical solutions for circular orbits in the \texttt{LogarithmicHaloPotential} at constant velocity, with a maximum deviation below 0.1\% after twice the characteristic time. Finally, we tested the case of a constant FDM factor against the analytical solution for circular orbits in the same potential at constant velocity, again finding a maximum deviation below 0.1\% after the characteristic time.

\end{document}